# A Big-Data based and process-oriented decision support system for traffic management

Alejandro Vera-Baquero[1,*] and Ricardo Colomo-Palacios[2]

[1]Universidad Carlos III de Madrid, Av. de la Universidad, 30, 28911 Leganés, Madrid, Spain.
[2]Østfold University College, B R A Veien 4, 1783 Halden, Norway.

## Abstract

Data analysis and monitoring of road networks in terms of reliability and performance are valuable but hard to achieve, especially when the analytical information has to be available to decision makers on time. The gathering and analysis of the observable facts can be used to infer knowledge about traffic congestion over time and gain insights into the roads safety. However, the continuous monitoring of live traffic information produces a vast amount of data that makes it difficult for business intelligence (BI) tools to generate metrics and key performance indicators (KPI) in nearly real-time. In order to overcome these limitations, we propose the application of a big-data based and process-centric approach that integrates with operational traffic information systems to give insights into the road network's efficiency. This paper demonstrates how the adoption of an existent process-oriented DSS solution with big-data support can be leveraged to monitor and analyse live traffic data on an acceptable response time basis.

Received on 22 February 2018, accepted on 12 May 2018, published on 29 May 2018

*Corresponding author. Email:averabaq@gmail.com

## 1. Introduction

As McKinsey forecasted in a report in 2014, the top priority objective of IT organizations at nowadays is generating greater value from data [1]. This objective lays among other IT advances on Big Data technologies. Not in vain, according to Forrester, other of the top players in the consultancy arena, Big data is a key trend for many industries [2]. Finally, Gartner, other of the world's leading information technology research and advisory companies, has been pointing out at Big Data as one of the strategic technology trends to watch in the last years. Moreover, Gartner states that Big Data is destined to help organizations drive innovation by gaining new and faster insight into their customers [3].

Scientific and commercial literature has provided many definitions for Big Data but agreeing with [4], it means data that is too big, too fast or too hard for existing tools to process. In any case, the fact is that due to the Data Tsunami that comes from the expanding mobile base, internet usage including social networks and sensors, naming just a subset of the data sources; managing and analyzing this information in an accurate way become a critical success factor for organizations around the world.

Big data phenomenon is affecting a wide range of fields of study. Thus, computer scientists, physicists, economists, mathematicians, political scientists, bio-informaticists, sociologists, and other scholars are clamoring for access to the massive quantities of information produced by and about people, things, and their interactions [5]. As a result of this, research driven by big data reflects a discipline that, to extract meaning from very large datasets, incorporates various techniques such as data mining and visualization into diverse fields [6]. However, the problem of Big Data is not new and the origin of big data research is rooted in various initiatives started in the early 1970s [6], being the book "Concise Survey of Computer Methods" [7] the most credited origin of the discipline. However, according to [8], the early 1990s is the beginning of the field of big data research. Whatever the case may be, although research based on big data can be conducted in various ways, its basic purpose lies in handling huge amounts of data from technological, sociological, and economic systems to discover some hidden patterns [6].

Smart cities and their infrastructures are one of the main producers of data and, as a consequence of this, Big Data efforts with regards to urban or traffic data are one of the most important arenas for research. The so called real-time city is a concept based on the use of real-time analytics to manage aspects of how a city functions and how it is regulated [9], in other words the key is using big-data enabled flow systems in order to enable the concept of smart city [10]. Initiatives include intelligent transportation systems based on mobile devices and big data architecture [11], a traffic temporal-spatial-historical correlation model based on big data [12], an infrastructure to support data management based on Hadoop HDFS and HBase as well as a set of web services to retrieve the data [13] or using the Open Source Data analytics platform Konstanz Information Miner (KNIME) to predict traffic distribution in the city of Santander in northern Spain [14]. However, in spite of the work performed by researchers around the world, and to the best of authors' knowledge, there is not an approach based on business process analytics and big data architecture to monitor and analyse the vehicles flows on roads for traffic management and control.

We present a big data based technological approach aimed to bring operational decision support technology to the traffic managers in order to help them drive traffic strategies on road networks.

## 2. Background

Large urban areas have undergone an exponential increase in the transport demand in the last decade, and congestion and roads safety have become a major concern to traffic managers and decision makers [15]. Besides, existing systems for traffic control are neither profitable nor efficient in terms of performance, cost, maintenance, and support [16], and normally these are based on estimations rather than on observed facts. With the advent of ubiquitous sensor data and cutting-edge technology such as Internet-of-Things (IoT), cloud-computing and big data, it could be possible to provide traffic managers with fact-based platforms that perform data-intensive operations in nearly real-time. In this regard, sensors can act as powerful data sources that are continuously collecting real-time traffic information for every single vehicle upon a certain period of time and relay this information to the processing units to infer its traffic parameters based on its trajectory. By making this extensible to every traceable vehicle in the area, it could be possible to determine the current state of any critical path at any time, in real-time and based on observable facts without the need of estimating traffic flows. Additionally, if that information is gathered and put together on a large enough scale, it could be used as input to advanced analytical methods to detect patterns or even predict future traffic behaviour. Moreover, if certain valuable analytical information is published and shared with citizens' in the form of a set of smarts services, then these services could be consumed by smart devices with the aim of helping local authorities bringing sustainability to the city by reducing congestion, improving the road safety and controlling $CO_2$ emission [17]. Nevertheless, the proliferation of IoT crowd-sourced data and sensors may arise privacy concerns about handling individuals' vehicle-tracing data on smart cities. Due to what mentioned above, there exists an increasing demand at nowadays for more advanced innovative systems that can adopt leading technology, not only to empower the functional capacity of existing systems, but also to improve the data-driven decision support of traffic managers and make cities smarter and more profitable [18] while preserving data privacy on cloud-computing networks and big-data contexts [31-34].

Smart technologies and Intelligent Transport Systems (ITS) are recently gaining relevance in the industry due to the latest advances in Information and Communication Technologies (ITC). ICT infrastructures play an important role in providing innovative smart services and novel applications in urban areas [19]. Pervasive ICT solutions encompass a wide range of technologies and ubiquitous services that can be applied to a city in multiple domains in general and traffic management in particular. In general, smart urban transport systems foster the adoption of smart technology which require cutting-edge communication and processing capabilities to deal with the heterogeneity nature of the data volume and speed processing demands [20]. Debnath et al. (2014) define smart technology as a self-operative and corrective system that consists of three elements: sensors, command and control unit (CCU), and actuators. These systems have the capabilities for sensing, communicating, acting and processing data for decision making [21]. In this regards, and more specifically in the context of intelligent traffic and transport management, data processing involves dealing with very large data streams to effectively identify and manage congestion issues and ensure the quality of service [18]. In turn, the monitoring and analysis of high volumes of data is a data-intensive process that usually exceeds the processing capabilities of a single large computer, thereby requiring robust and complex supporting systems that can easily scale over time to meet the processing demands in terms of latency and data volume. Hence, next generation of traffic management systems must be big-data ready for enabling elastic-scalable data analysis.

Over the recent years, the latest advances in sensing and communication technologies have attracted researchers'

attention from both industry and academia as a mean to improve the existing road traffic management systems [22]. However, more efforts must be pursued in order to produce scalable analytical systems that can deal with the heterogeneity issues and variety and volume complexities of processing crowd-sourced streams [18] [23] [24] [25]. Cloud-based solutions for providing scalable traffic management systems have been discussed and proposed in [26] and [27] respectively, but these are insufficient for completing the analytical demands cycle aforementioned. Therefore, further research efforts are still needed to address the complexities and challenges aforementioned for bringing fully big-data operational support to traffic management systems.

The authors propose a process-centric solution that leverages recent advances in process performance management on big-data [28] contexts with the aim of applying its fundamentals to the traffic management realm.

## 3. The Approach

We propose the adoption of a former work in the area of business process analytics to monitor and analyse the vehicles flows on roads for traffic management and control. The proposed analytical framework is widely discussed in [29] and has the ability to provide cloud computing services in a timely fashion. These services rely on a big-data infrastructure that enables the system to perform data-intensive computing on processes whose executions produce a vast amount of event data that cannot be efficiently processed by means of traditional systems. This data usually comes from a variety of heterogeneous platforms that are continuously producing enterprise business information.

The problem of integrating event data from heterogeneous data sources data can be challenging and is very common to occur on distributed environments where the continuous execution of distributed processes produces high volumes of unstructured event data that may occur on a variety of diverse platforms. The framework overcomes this pitfall by introducing a generic event model that is an extension of BPAF (Business Process Analytics Format) [30], hereinafter exBPAF. This format enables heterogeneous systems to produce interchangeable event information regardless of the underlying concerns of the source systems, and it provides the information required to enable the DSS to perform analytical processes over them, as well as representing any derived measurement produced during the execution of any process flow. This is essential to provide the framework of a concrete understanding and representation of what needs to be monitored, measured and analysed, and it can contribute to the continuous improvement of processes by giving insights into process performance.

This framework is also referred in the literature, as Decision Support Systems (DSS) [29] [35], which have been used in several case studies to perform business process analysis in the business process improvement discipline. In addition, this solution has the capabilities to timely collect the enterprise events, correlate the data along their inter-related processes, and measure their throughput. These extraordinary capabilities could be adopted by the industry to track every single vehicle traveling along the road network and monitor live traffic data flow on a short response time basis. Likewise, this can lay the ground for establishing a fact-based analysis over the gather data instead of using idealized models or estimation basis. Thus, it may enable analysts to gain insights into the past, present and future of traffic behaviour by mining very large data repositories. Hence, we propose to leverage this process-centric DSS and apply its fundamentals to the traffic management realm with the aim of analysing live traffic data

### 3.1. A Traffic Process Model

For adopting the business process improvement principles into the traffic management discipline, we need to link these two areas of knowledge by establishing a representation of the traffic flow as a process. For this purpose, we use a generalized representation of a road network based on a graph definition $G = (J, S, R)$ where $J$ are the vertices (junctions), $S$ are the edges (road sections), and $R$ are the roads. Let $J$ be the set of junctions $J(G) = \{j_1, j_2, j_3, ..., j_n\}$, let $S$ be the set of road sections $S(G) = \{s_1, s_2, s_3, ..., s_n\}$, and let $R$ be the entire set of roads in the network $R(G) = \{r_1, r_2, r_3, ..., r_n\}$. We denote $R(i) \subseteq S$ as the set of sections of the road $i$, so that every road section must correspond to only one road $\forall i, j \in R : i \neq j \rightarrow R(i) \cap R(j) = 0$. Similarly, we define a path, denoted by the function $P(x, y) \subseteq S$, as the sequence of adjacent edges that link the source **x** to the destination **y**. The next figure illustrates an example of a road network definition by following these premises.

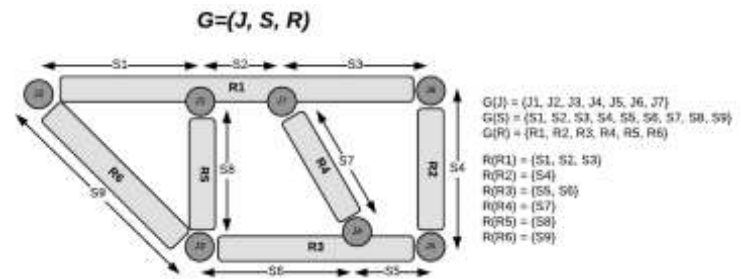

**Figure 1**. Sample of road network modelling

The time function $T(s, v, t)$ represents the time that takes an individual vehicle $v$ to drive along the road section $s \in S$ in a particular period of time $t$. Consequently, the time consumed by a vehicle to drive along a full path is given by the sum of its individual times. This is formulated by the following equation:

$$T(P(x,y),v,t) = \sum_{s \in P(x,y)} T(s,v,t)$$

The flow function $F(s,t)$ represents the number of vehicles that drive along the road section $s \in S$ in a particular period of time $t$. Consequently, the flow on a path is determined by the sum of its individual flows. This is formulated by the following equation:

$$F(P(x,y),t) = \sum_{s \in P(x,y)} F(s,t)$$

The interest of the study rely on the identification of bottlenecks on different paths, as this will enable traffic managers to gain a broader vision of the state of road network at any time, and thus performing congestion analysis in nearly real-time. The vector function $\vec{F}_p$ is an analytical measure that gauges the flow of individual sections on a given path $P(x,y)$ for a particular period of time $t$, and it is formulated by the following equation:

$$\vec{F}_p(P(x,y),t) = \langle F(k_1,t), F(k_2,t),...,F(k_n,t) \rangle, k_i \in P(x,y)$$

Likewise, the vector function $\vec{T}_p$ is an analytical measure that specifies the time that it takes a vehicle to travel along individual sections on a given path $P(x,y)$ for a particular period of time $t$, and it is formulated by the following equation:

$$\vec{T}_p(P(x,y),v,t) = \langle T(k_1,v,t),...,T(k_n,v,t) \rangle, k_i \in P(x,y)$$

From a process perspective, the journey of a vehicle can be modelled as a process whose activities represent the set of adjacent edges $S(G)$ that a vehicle travels through to reach its destination. Thereby, we represent a process model as a specific path in the road network that connects a source *x* with a destination *y*. This is denoted by $p_{x,y}^{def} = P(x,y) \subseteq S(G)$. In addition, we model a process instance as a running vehicle that is traveling through a specific path upon a certain period of time *t*, and this is denoted by the pairs $(v,t)$. Similarly, such path is represented by the process model $p_{x,y}^{def}$. The set of adjacent edges correspond to the number of inter-related and consecutive activities that are specified in the vehicle's journey (process definition), and the completion time of the activities represent the time a particular vehicle *v* takes to drive along the road sections *s* upon a particular time *t*, and this is denoted by the ternary $(s,v,t)$.

Within a process context, let's define the following:

- $p_{x,y}^{def} \equiv$ *Process definition for the path* $P(x,y)$
- $p \equiv$ *Process instance of the vehicle* ***v*** *traveling through the path* $P(x,y)$ *on a particular period of time* ***t***
- $a^{def} \equiv$ *Activity definition of the process instance* ***p*** *that is traveling through the road section* $s \in S(G)$
- $a \equiv$ *Activity instance of the vehicle* ***v*** *traveling through the road section* ***s*** *on a particular period of time* ***t***

The following table summarizes the analogy aforementioned between a traffic-oriented and process-centric model.

**Table 1**. Traffic process model definitions

|  | Traffic-oriented model | Process-centric model | Attributes |
|---|---|---|---|
| **Process definition** ($p_{x,y}^{def}$) | $P(x,y) \subseteq S(G)$ | $p_{x,y}^{def} = P(x,y)$ | $p_{x,y}^{def} \equiv$ Model definition of the process |
| **Process instance** ($p$) | $(v,t)$ | $p_{model} = P(x,y)$ $p_{id} = v$ $p_{time} = t$ | $p_{model} \equiv$ Model of the process instance ***p*** $p_{id} \equiv$ Identifier of the instance ***p*** $p_{time} \equiv$ Time of the process instance ***p*** |
| **Activity definition** ($a^{def}$) | $s \in S(G)$ | $a^{def} = s$ | $a^{def} \equiv$ Model definition of the activity |
| **Activity instance** ($a$) | $(s,v,t)$ | $a_{model} = s$ $a_p = p$ $(a_p)_{id} = v$ $a_{time} = t$ | $a_{model} \equiv$ Model of the activity ***a*** $a_p \equiv$ Process instance of the activity ***a*** $a_{id} \equiv$ Identifier of the activity ***a*** $a_{time} \equiv$ Time of the activity ***a*** |

The traffic-process model that we aim to monitor and analyse is very basic and simple (see Figure 2). The journey process is initiated upon vehicle detection on the road network. Once the system has knowledge of the existence of a particular vehicle, its journey is recorded by tracking the plate registration number on every junction, thereby the road sections are identified as the vehicles progress along its journey. The process ends either when the vehicle goes beyond the limits of the scoped area, or it reaches its destination.

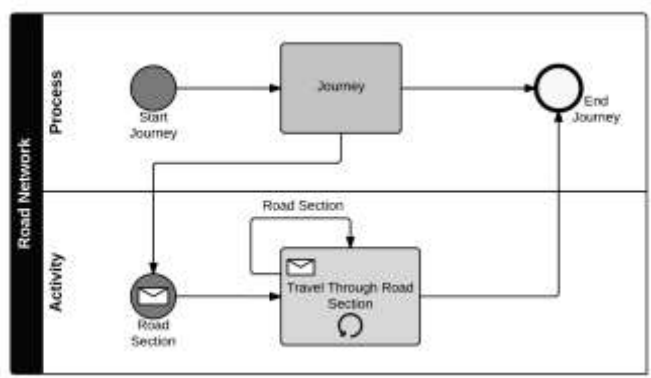

**Figure 2**. Traffic-process model

## 3.2. The Analytical Framework

As previously mentioned, the analytical framework proposed is based on a process-oriented DSS solution presented in [29] which is focused on the improvement of process performance by analysing the execution outcomes of the processes. One of the major contributions of this work relies on the ability of the framework to be agnostic to any specific domain. In this regard, we aim to exploit this capability by bringing its analytical features into the traffic management realm.

### 3.2.1. The Architecture
The DSS proposed is a cloud-computing and big-data based solution that supports distributed process analysis and provides an analytical environment to business users for performing timely data analysis. The overall architecture of the framework is depicted on Figure 3. It consists of a variety of cloud-based analytical services that are integrated into a main functional unit so-called GBAS (Global Business Analytics Service). This unit is the key component responsible for providing monitoring and analysis functionality to third-party applications. Likewise, a number of BASU (Business Analytics Service Unit) nodes are provisioned along the collaborative network for enabling local analysis within a specific area. Every single BASU unit supports the monitoring and analysis of the data coming from its area of scope, thus it provides the capabilities to capture and integrate heterogeneous event data that is originated from diverse traffic operational systems. Those units are attached to one or many areas of the road network and they handle their own big-data based repository in isolation with respect to the rest of BASU units. The number of nodes may vary depending on two factors: 1) the volume of data, and 2) the performance of the DSS. Obviously, these two factors are related to each other, but it is important to analyse them individually when setting up the environment. As the volume of data grows, the computational resources needed to keep latency in low levels increase, and thus the number of nodes determines the trade-off between hardware investments and low response rates.

The integration of BASU subsystems is required for measuring the traffic performance of cross-functional areas that are extended beyond the software boundaries. The GBAS (Global Business Analytics Service) module is the entity responsible for integrating BASU units along the entire road network and providing analytical insights into traffic performance of inter-related areas. Consequently, a large number of ANPR publishers are attached to the BASU units to feed the nodes with timing information about vehicles flowing through road sections. Moreover, there must be at least one ANPR system settled in every junction, and these publishers are continuously listening to its operational ANPR systems for incoming registration numbers of vehicles. Once the vehicle is identified, the publishers send timing event information to the analytical units for processing. The integration of BASU subsystems is required for measuring the traffic performance of cross-functional areas that are extended beyond the software boundaries. The GBAS (Global Business Analytics Service) module is the entity responsible for integrating BASU units along the entire road network and providing analytical insights into traffic performance of inter-related areas. Moreover, this key component is the core point for providing analytical services to third-party applications. In a nutshell, the cloud-service components (BASU & GBAS) have the capabilities for collecting data originated from distributed heterogeneous traffic systems, storing the live traffic data by using big data underlying technology and inferring knowledge from the gathered information. This analytical framework has the ability to store and process massive amount of vehicles journeys over time. This allows traffic managers to not only determine the exact time and location of individuals on the fly, but also to analyse historical snapshots of conflictive roads, incorporate behavioural pattern recognition on hot-spots or congested roads, and even predict future behaviours by leveraging predictive analysis methods and simulation tools.

### 3.2.2. Traffic Systems Integration
Current technologies such as Closed-Circuit Television (CCTV) and Automatic Number Plate Recognition (ANPR) can play an important role in identifying traffic movement and location based on vehicle registration plate numbers. ANPR systems are widely used in many urban areas and productively adopted across Europe. These systems have been included in the architectural solution with the aim at feeding the system with traffic status information.

Vera-Baquero, A. & Colomo-Palacios, R.

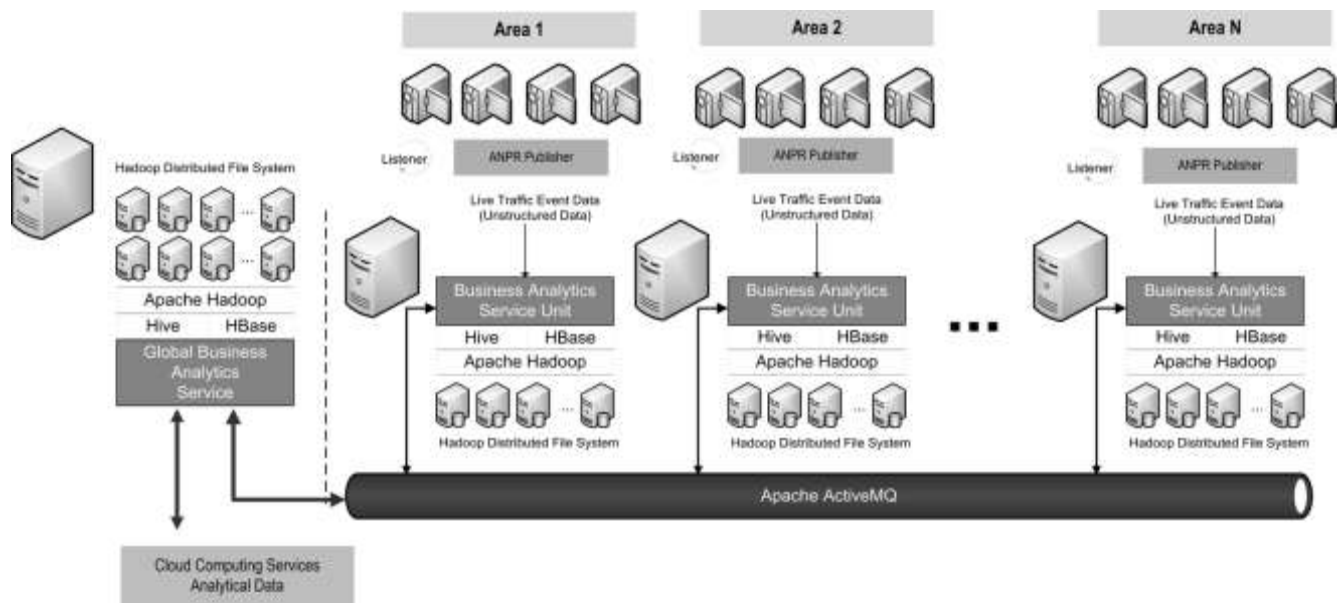

**Figure 3.** Architecture of the analytical framework

Traffic operational platforms, namely ANPR systems, are intended to produce unstructured event data which contains the captured vehicle registration number, the exact date time, the area on where the camera is located and the ANPR camera identification itself. These cameras are located in every single junction of the scoped area we aim to monitor and control.

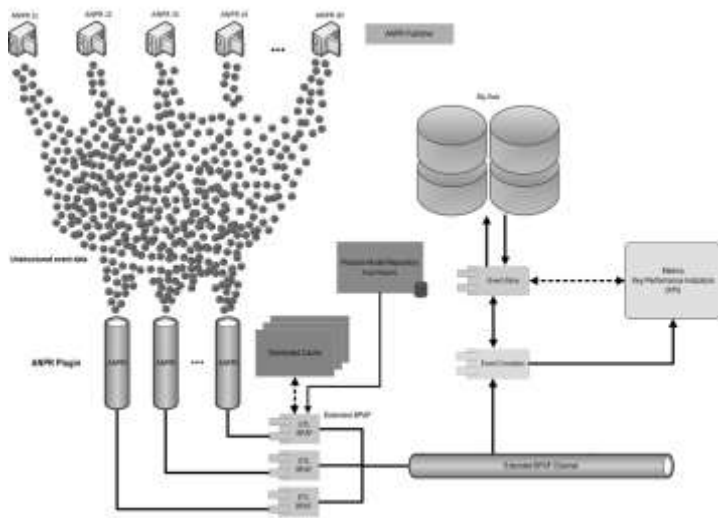

**Figure 4.** Traffic live event data processing

The Figure 4 illustrates the event-capturing side of the proposed IT solution. Multiple ANPR plugins receive these t-uples from the software listener and generate in parallel a massive amount of events in BPAF (Business Process Analytics Format) standard format [32]. These events are then forwarded to an attached processing intermediate module which converts the source event streams into exBPAF (extended BPAF) format [33] by the means of specific ETL (Extract, Transform & Load) methods. The ultimate goal of this module basically consists in discovering the road section the vehicle is travelling trough. This is achieved by querying the vehicle's junction record over the last known trajectory. This information is temporarily stored in an internal distributed cache with data eviction and is matched against the process model repository, which represents the road network map we are analysing and monitoring. Once the road section (activity) is identified, the event data is then emitted and published to the network throughout the ActiveMQ message broker, which will forward the generated event to the internal BASU dependencies for processing. At this stage, the events contain all information needed to infer the process instance of the vehicle at hand.

### 3.2.3. Event-Based Model

BPAF is a standard format published by the Workflow Management Coalition to support the analysis of audit data across heterogeneous process-aware systems. It enables the delivery of basic frequency and timing information to decision makers, such as the cycle times of processes, wait time, number of process instances completed against the failed ones, etc. This permits host systems to determine what has occurred in the business operations by enabling the collection of audit data to be utilized in both analysis and the derivation of status information [30]. An event in exBPAF [33] represents a state change accordingly to the transitions specified in the BPAF standard, and consists of at least a unique event identifier, the identifiers of the process definition and instance (if available), a timestamp, some event correlation information and a state transition [29]. Within the context of this work, this standard allows the system to audit vehicle journeys and derive measurement information from the event sequence. This is achieved by tracking the state transitions over the instance lifetime. The next table outlines the most relevant exBPAF fields related to the study.

**Table 2**. exBPAF description

| Attribute | Description | Model |
|---|---|---|
| **EventID** | A globally unique identifier for the individual event. | $e_{id}$ |
| **Timestamp** | The time of the event occurrence. | $e_{time}$ |
| **ServerID** | A globally unique identifier for the source ANPR system of the event. | $e_{srcid}$ |
| **Process DefinitionID** | The identifier of the journey the vehicle instance is flowing through. | $e_{pdef}$ |
| **Process InstanceID** | The identifier of the vehicle instance for the event journey. | $e_{pid}$ |
| **ProcessName** | The name of the journey. | $e_{pname}$ |
| **Activity DefinitionID** | The road section. | $e_{adef}$ |
| **Activity InstanceID** | The identifier of the vehicle instance travelling through the road section. | $e_{aid}$ |
| **ActivityName** | The name of the road section. | $e_{aname}$ |
| **CurrentState** | Current BPAF state transition. In this model there are only two possible values: OPEN_RUNNING and OPEN_CLOSED_COMPLETED. | $e_{current.state}$ |
| **PreviousState** | The previous BPAF state. In this model there is only one possible value: OPEN_RUNNING or NULL for the starting instances. | $e_{previous.state}$ |
| **Correlation** | A name-value-pair that stores a subset of elements that are used to uniquely identify a determined process or activity. | $e_{correlation=\{reg,date\}}$ Reg: Vehicle registration number. Date: Journey date. |

The system captures and records the timestamp of events containing the time at which they occurred on the ANPR system, not when they are packaged or delivered. This property is essential in order to identify and analyse the correct sequence of activity instances, as well as ensuring that the generation of metrics produces precise information on its outcomes. The vehicle path is reconstructed by retrieving the sequence of all events for a particular instance, and in turn, the path definition base is generated progressively as the events arrive.

The exBPAF event model permits the representation of the structural and behavioural information, but this is not enough to provide traffic managers with timely performance information. The correct sequence of instances (vehicle journeys) must be identified before any measurement can be applied, and thus, this action must be executed in a very-low latency. The event correlation mechanism is based on the shared data between instances during their execution. This information makes reference to the payload of event messages, and this is used to identify the start and end event data for a particular vehicle journey. In this case, and assuming a road network analysis on a daily basis, the correlated data in this configuration are the vehicle registration number and the journey date. With this information, the system is able to identify the journey of a particular vehicle on a given time. The event correlation algorithm is out of scope in the present paper, but its specific details are fully expanded and widely covered in [28]. Whilst the live event traffic data give an insight into the vehicle journeys, they do not provide measurable information about road network performance, therefore the definition of metrics are required in order to provide analysts with an understanding of the behavioural aspects of the data. Hence, the metrics and key performance indicators (KPIs) are essential to build a concrete understanding of what needs to be monitored and analysed [33].

### 3.2.4. Key Performance Indicators

Within a real-time monitoring context, the construction of metrics and KPIs is intended to be performed at minimum latency in order to enable traffic managers to react quickly to undesired situations, and this is a data-intensive process in the DSS [31]. To this extent, this is similar to what happens on BAM (Business Activity Monitoring) solutions, where the business activity outcomes must be timely accessible to business users for decision making [28]. Therefore, the BAM-like component of the DSS plays an important role in the KPIs configuration. This feature is very useful to managers as they may decide whether or not establish thresholds per process (path) or activity definitions (road sections). This depends on whether there already exists historical information in the DSS, as this will allow the system to infer the expected execution time of a process or activity. In such case, the thresholds might be set in the BAM component to generate alerts for detecting bottlenecks and congestion in nearly real-time. However, the activation of these thresholds entails the timely generation of metrics per vehicle along with its aggregations, and this is a computing-intensive operation whereas the system is tracking individual vehicles over time. In this regard, the volume of data is presumably to be high, depending on the flow rates of the road network, and thus, the activation of these alerts must be selected with caution. Furthermore, they should be deactivated if they are not essential for the purpose of the analysis, namely when performing historical or predictive analysis.

In order to achieve a timely generation of metrics, the immediate events are temporarily stored in a distributed cache with data eviction. Since the events are correlated as they arrive, the path sequence of the vehicles co-exists during a variable period of time, in both permanent storage and cache, and its metrics and aggregated KPIs are calculated and updated accordingly once an event transition towards a COMPLETED state, either for an activity (road section) or process (journey) instance. At that precise moment, the entire event stream is read from cache for that particular instance in place and forwarded to the data warehouse module for processing. Then the entire event stream sequence for a single instance is analysed and the metrics are produced according to the state transition changes based on zur Muehlen & Shapiro's model [32]. The event cache has been implemented using Infinispan and configured as a distributed cache with replication (owner nodes). Further detailed information can be found at [28].

At following we outline some metrics that the DSS can deal with, but not limited to:

**Processing time**: Measures the net execution time per process instance or activity. In the case of process instances, it refers to the global journey time of the vehicle. By contrast, for activity instances it refers to the time a vehicle takes to travel along a road section.

**Activity Instance**
The processing time (T) of a determined vehicle travelling through a specific road section ($a_{id}$) is given by the difference of time between the start and end events for that particular instance.

*Time metrics*
$$T(a_{id}) = e_{time}^{CLOSED.COMPLETE} - e_{time}^{OPEN.RUNNING} : e \in E / e_{aid} = a_{id}$$

*Flow metrics*
The flow metric (F) of a determined vehicle in a road section is 1.

$$F(a_{id}) = 1$$

**Activity Definition**
The average processing time (T) of a determined road section ($a_{model}$) upon a certain period of time $\Delta t \equiv [t_1, t_2]$ is given by the sum of the processing time of all its activities divided by the total number of instances.

*Time metrics*
$$T_{AVG}(a_{model}) = \sum_{e \in E / e_{adef} = a_{model} \wedge e_{time} in.\Delta t} \frac{T(e_{aid})}{count(e_{aid})}$$

The flow metric (F) of a determined road section upon a certain period of time $\Delta t \equiv [t_1, t_2]$ is given by the number of vehicles (activities) that are travelling or have travelled through that road section ($a_{model}$). This is determined by its number of events which have transitioned to state OPEN_RUNNING within that interval of time $\Delta t$.

*Flow metrics*
$$F(a_{model}) = \sum_{e \in E / e_{adef} = a_{model} \wedge e_{time} in\, \Delta t \wedge e_{OPEN.RUNNING}} 1$$

**Process Instance**
The processing time (T) of a determined vehicle travelling through a specific path ($p_{id}$) is a vector metric that comprises the individual processing time of every road section along the path.

*Time metrics*
$$\vec{T}(p_{id}) = \langle T(e_{aid})_1, T(e_{aid})_2, ..., T(e_{aid})_n \rangle : e \in E / e_{pid} = p_{id}$$

The flow metric (F) of a determined journey for a particular vehicle is a vector metric that encompasses the individual flows along the road sections.

*Flow metrics*
$$\vec{F}(p_{id}) = \langle F(e_{aid})_1, F(e_{aid})_2, ..., F(e_{aid})_n \rangle : e \in E / e_{pid} = p_{id}$$

**Process Definition**
The average processing time (T) of a specific journey ($p_{model}$) upon a certain period of time $\Delta t \equiv [t_1, t_2]$ is given by the sum of the processing time of all its processes divided by the total number of instances in that journey.

*Time metrics*

$$T_{AVG}(p_{model}) = \sum_{e \in E / e_{pdef} = p_{model} \wedge e_{time} in.\Delta t} \frac{T(e_{pid})}{count(e_{pid})}$$

The flow metric (F) of a determined journey ($p_{model}$) upon a certain period of time $\Delta t \equiv [t_1, t_2]$ is a vector metric that encompasses the individual flows along the road sections for that journey within that interval of time $\Delta t$.

*Flow metrics*

$$\vec{F}(p_{model}) = \langle F(e_{adef})_1, F(e_{adef})_2, ..., F(e_{adef})_n \rangle :$$
$$e \in E / e_{pdef} = p_{model} \wedge e_{time} \ in \ \Delta t$$

## 4. Case Study: UK road network

We present a case study that aims to demonstrate the feasibility of applying the approach proposed in this paper to the traffic management realm. The adoption of the big data based DSS presented in [29] along with the traffic model introduced in the previous section is leveraged to analyse the efficiency of the roads network in England. The study is focused on the motorways and major trunk roads, as these are the roads with highest traffic flow rates, and are the most interesting cases to identify hot-spots and congestion. As part of the evaluation, we have used a real-life data set published by the Highways Agency, which is publicly available and fully accessible in [34]. This dataset provides average speed, journey times and traffic flow information on all motorways and 'A' roads. This is known as the Strategic Road Network in England.

The setup of the DSS is performed by following the methodology discussed in [31]. During the define phase we identified and represented the model illustrated in Figure 2, and determined the scope and boundaries of the study. Due to managerial and performance reasons, we have broken down the analysis into six different areas: North West, North East, Midlands, East, South West and South East. The reason for splitting the data analysis into different regions is twofold: 1) managerial reasons: the analysis is greatly simplified when it is performed per areas as it allows traffic managers to perform data analysis locally on specific locations rather than dealing with a broader view of the entire road network, and 2) performance reasons: the high volumes of live traffic data are easier to manage when they are stored and analysed in isolation. Likewise, we determined the process models by selecting a variety of journeys that connect two cities through different routes. For doing this, we constructed the model table based on a road map and set the routes of every journey along with the properties of every road link (road section). This information is supplied in the data set (refer to the data set published information in [34] for further details). The following table outlines a sample of one the process models developed.

**Table 3**. Process model sample

| Process | Activity | Properties* |
|---|---|---|
| **BirmStaf01** #Journey [Birmingham-Staffordshire] | LM1012 Start Junction: M6 J6 End Junction: M6 J5 | PN, JD, TP, FR, AS, LL, LD |
| | LM1011 Start Junction: M6 J5 End Junction: M6 J4A | PN, JD, TP, FR, AS, LL, LD |
| | LM513A Start Junction: M6 J4A End Junction: M42 J8 | PN, JD, TP, FR, AS, LL, LD |
| | LM1052A Start Junction: M42 J8 End Junction: M6 T11 | PN, JD, TP, FR, AS, LL, LD |
| | … | |
| **BirmStaf02** #Journey [Birmingham-Staffordshire] | LM1015 Start Junction: M6 J6 End Junction: M6 J7 | PN, JD, TP, FR, AS, LL, LD |
| | LM1017 Start Junction: M6 J7 End Junction: M6 J8 | PN, JD, TP, FR, AS, LL, LD |
| | LM1019 Start Junction: M6 J8 End Junction: M6 J9 | PN, JD, TP, FR, AS, LL, LD |
| | LM1021 Start Junction: M6 J9 End Junction: M6 J10 | PN, JD, TP, FR, AS, LL, LD |
| | … | |

*PN=Plate Number, JD=Journey Date, TP=Time Period, FR=Flow Rate, AS=Average Speed, LL=Link Length, LD=Link Description*

In the configuration phase we provisioned six analytical nodes (BASU) attached to every single area, thereby each node manages its own data in isolation with respect to the rest of nodes. These units are BASU-SW (South West), BASU-SE (South East), BASU-EA (East), BASU-ML (Midlands), BASU-NW (North West), BASU-NE (North East). Whilst BASU nodes provides local analysis within a single area, the GBAS node provides inter-regional analysis by managing traffic data that represent flows that pass through roads that link different areas. After the provision of nodes, we loaded the process models (journeys) into the units and we proceeded to implement the listeners. In this regard, we simulated an Automatic Number Plate Recognition (ANPR) system that consisted in producing event streams that modelled the behaviour of the real traffic operation system described herein. The simulator software generated pseudo-random numbers from a normal distribution based on the values provided in the dataset. The plate registration numbers were randomly created and the journey time and traffic flow (number of vehicles per road) were inferred from real information publicly available.

| Activity (Road Section) | Dataset mean | Observed mean | Standard deviation | t-score | p-value | Level of confidence |
|---|---|---|---|---|---|---|
| LM1012 | 224.32 | 224.2510299 | 11.57346702 | 0.171479701 | 0.863888539 | 0.01 |
| LM1011 | 152.93 | 153.082836 | 7.46496521 | 0.655800196 | 0.512099861 | 0.01 |
| LM513A | 36.07 | 36.07375657 | 1.833118233 | 0.03285243 | 0.973817867 | 0.01 |
| LM1052A | 37.13 | 37.09007038 | 1.844114323 | 0.693217078 | 0.488330529 | 0.01 |
| LM1047A | 37.95 | 37.69565383 | 2.106296919 | 2.088051481 | 0.037642039 | 0.01 |
| LM1045A | 83.93 | 83.89237389 | 4.283367395 | 0.154911682 | 0.876991756 | 0.01 |
| LM1042A | 241.88 | 242.0090013 | 12.7299671 | 0.17107606 | 0.864285782 | 0.01 |
| LM1040A | 40.06 | 39.99760023 | 2.147345071 | 0.489711508 | 0.624717442 | 0.01 |
| LM1037A | 180.17 | 180.5327666 | 9.879086223 | 0.654823239 | 0.513056415 | 0.01 |
| LM1036A | 215.23 | 215.0733824 | 11.20418408 | 0.236810675 | 0.812973196 | 0.01 |
| LM1034A | 78.93 | 78.69303597 | 3.740780291 | 1.091687797 | 0.275858132 | 0.01 |
| LM1033A | 36.47 | 36.57435774 | 1.919742799 | 0.833327196 | 0.405510001 | 0.01 |
| LM928B | 75.03 | 74.91076678 | 3.761801357 | 1.028526729 | 0.303938541 | 0.01 |
| AL3268 | 112.63 | 112.8836463 | 5.730868816 | 0.571960875 | 0.568122197 | 0.01 |

Similarly, the events reproduced the journey according to the model specified in the previous phase by fulfilling with the flow rate and time information of every road section for the given date. The information provided by this simulator was the vehicle plate number, the simulated date time, the designated area and the junction identifier that indicates on where the camera was located.

Once the publisher side of the framework was completed, we developed a specific plugin for processing the incoming events and convert raw APNR data into exBPAF. This plugin incorporates a specific ETL module for deriving the vehicle journey by querying the start and end junctions of consecutive events. Basically, it searches for the activity definition in place associated to the instance thereof.

**Functional Analysis**

The evaluation was performed in a test environment that meets the infrastructure depicted on the Figure 3. A vast amount of event data was generated by simulating the traffic flow experienced during May-Jun of 2013 in two different routes from Birmingham to Staffordshire, and we successfully experienced that the outcomes of the DSS were those expected. The execution results, measures and KPIs did not present any statistical significance with respect to the official values publicly available. In this regard, we used paired t-test with the aim of verifying that there was no significant difference between the two population means, i.e. the mean provided by the data set ($\mu_1$) and the one inferred by the system ($\mu_2$). We must bear in mind that the DSS calculates the execution time of processes by detecting state transitions on the sequence of events received from the capturing software. Any misidentification, inaccuracy or failure on the correlation algorithm would produce untrustworthy results, which in turn would necessarily entail a deviation between means. In order to verify that there is no statistical significance between the dataset and the DSS outcomes we carried out a paired t-test.

First, we define the null hypothesis (Ho) and the alternate hypothesis (Ha) such as follows:

Ho ($\mu_1 - \mu_2 = 0$) : *The DSS detects the real process behaviour (the system is accurate)*

Ha ($\mu_1 - \mu_2 \neq 0$) : *The DSS produce wrong outcomes (the system is not accurate)*

The null hypothesis states that the system reproduces the expected behaviour of processes, namely it conforms to those specified in the dataset. Thus we assert that the system is accurate and reliable, and we want to test whether this assertion is true or not by invalidating the null hypothesis (Ho). In order to do this, we collected the data generated by the DSS on the first 15 minutes interval for a specific date. Then we carried out a t-test per activity (road-link) by verifying that the p-value was greater than the level of confidence (0.01) for every road link. Upon test completion we could not reject the null hypothesis, thereby demonstrating that there is no evidence that lead us to think that system is inaccurate. Thus we conclude that the experiment showed that the DSS correlated instances and generated the process execution time metrics correctly with a 99% of confidence. The table above summarizes the results of every t-test undertaken per activity.

The comparative of results between the real data published in the dataset and the DSS outcomes for the route 1 is depicted on Figure. The figure illustrates the average time for vehicle to travel through the road links stated on the x-axis, whereas the y-axis displays the average journey time in seconds. It worth to point out that the dataset specifies the estimated journey time on a road link for a given date and within a 15 minutes interval, and the simulator produced the data to feed the DSS based on these estimations.

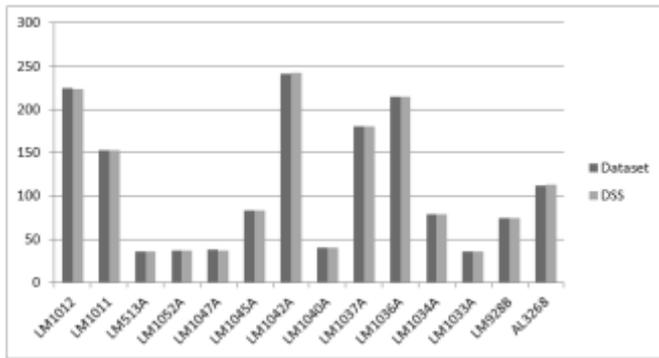

**Figure 5**. Comparative outcomes between the DSS output and real dataset

**Performance Analysis**

In regards of the system performance analysis, we carried out the evaluation of the framework by deploying the infrastructure on a Cloudera (CDH 4.4) instance using a 4-node configuration, thereby exploiting the clustering capabilities of the proposed solution. The volume of data for the route 1 rose to nearly 24 million vehicles, and the correlation algorithm performed reading operations in the order of few milliseconds for such volume of data. The simulation entailed the generation of two events per vehicle for every road link the vehicle is travelling through. This corresponds to approximately 640GB of raw data assuming a size of 1KB per event generated at source. It is important to highlight that this is a prospective study and further efforts are ongoing to progressively increase the volume of data to the levels of TBs of information by adding new routes to the model. As already sated, the most significant finding is that the read operations performed in the order of few milliseconds (see Table 4). This implies the event correlation algorithm can run at minimum latency, thus linking events as they occur without undergoing any delay that may impact the generation of metrics in nearly real-time. The second finding is that the read execution time remained stable over time and did not increase as the number of events grew. It is worth to point out that the system correlates the events as they arrive by finding their predecessors in the big data storage, whereby this entails one read operation plus a write for storing the event in the repository in order that it can be found by its successors. Once the events are correlated, these remain in memory for a given period of time in order to enable low-cost time access to sequence of events for the metrics generation (see 3.2.4). This is especially important for offering the system with real-time BAM capabilities. The next table shows that the IT solution features a high performance and is able to produce timely metrics by monitoring around 1300 vehicles per second using a small cluster size.

**Table 4**. Performance Analysis

|  | I/O ops | Average (ms) | Stdev (ms) | Throughput (events / sec) |
|---|---|---|---|---|
| **Event Corr.** | Read | 0.299694 | 1.779816 | 3337 |
|  | Write | 0.465463 | 1.815193 | 2148 |
| **Metric Gen.** | 1 read + 1 write | 0.765157 | 1.797592 | 1307 |

Future endeavours will be focused on extending the battery of tests to incorporate extra routes, sources and destinations, as this will make the data grow considerably. This will put the research work in the right direction to determine the trade-off between volume of data, KPI latency and hardware investments.

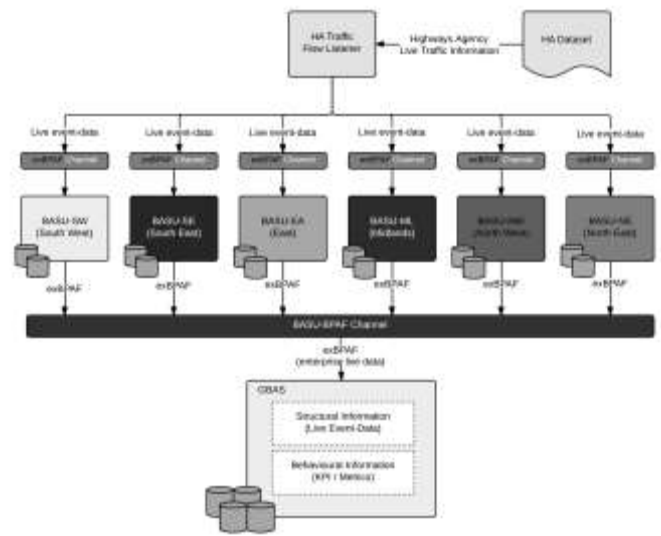

**Figure 6**. DSS infrastructure

## 5. Conclusions and Future work

The application of a former work in the field of business process analytics has been proposed with the aim at enabling timely analysis of massive vehicle data flows on road networks. The outcome of this former work is a process-oriented DSS system that enables business users to analyse and improve distributed business processes that are run on heterogeneous platforms. This approach has been harnessed and brought to the traffic management field for enabling the monitoring and analysis of very large amount of data in nearly real time. This has been attained by the adoption of big data technology, which is the key enabler for tracking the movement of single vehicles on a road network over time. This offers plenty of possibilities for exploiting the gathered information and performing traffic analysis based on the observable facts. By doing this, it could be possible to identify quickly the location of stolen vehicles, compare different routes between junctions for a particular period of time with the aim at redistributing the traffic flow, applying what-if simulations over different routes, or even use behavioural pattern recognition for the automatic detection of conflictive and congested roads.

The downside of this approach is that the solution proposed highly depends on the reliability of the ANPR source systems. Any failure or miss at the plate number detection may generate wrong paths for such vehicle, thus creating inconsistencies in the big data repository. Those, cases should be rejected by the system in order to keep the analytical data as trustworthy as possible. However, as already stated, our system relies on the facts not on the models, and whatever process execution path is acceptable. That means that the instances do not have to fulfil the normative model, and in contrast, the model is reconstructed from the processes execution itself. The current study is a special case as there will never be deviations in respect with the normative model, and thus processes must always comply with the road network definition. Otherwise, non-compliant processes must be discarded. Such functionality is not supported by the framework yet and is part of future work. In this regard, we also aim to gradually incorporate new advanced analytical functionality, such as applicable process mining techniques, predictive analysis or event stream reproduction in simulation mode.